\documentclass{article}
\usepackage{mypreamble}
\usepackage{titlesec}
\usepackage{multirow}
\usepackage{appendix}
\usepackage{chngcntr}
\usepackage{subcaption}

\usepackage[backend=biber, style=apa]{biblatex}
\usepackage{tikz}
\usepackage{geometry}

\usetikzlibrary{positioning, calc}
\usetikzlibrary{decorations.pathreplacing}
\usetikzlibrary{patterns,patterns.meta}
\usetikzlibrary{external}
\tikzset{circle node/.style = {circle,inner sep=1pt,draw, fill=white},
        X node/.style = {fill=white, inner sep=1pt},
        dot node/.style = {circle, draw, inner sep=5pt}
}
\definecolor{myblue}{RGB}{0,95,115}
\definecolor{myred}{RGB}{155,34,38}

\title{Bivariate deconvolution for cancer detection after surgery}
\author[1,$\ast$]{Nuria Senar}
\author[2,4]{Stavros Makrodimitris}
\author[1]{Michel H. Hof}
\author[3]{Cornelis Verhoef}
\author[4]{Saskia M. Wilting}
\author[1]{Mark A. van de Wiel}

\affil[1]{Epidemiology \& Data Science, Amsterdam UMC}
\affil[2]{Swammerdam Institute for Life Sciences, Universiteit van Amsterdam}
\affil[4]{Department of Medical Oncology, Erasmus MC Cancer Institute, UMC Rotterdam}
\affil[3]{Department of Surgical Oncology and Gastrointestinal Surgery, Erasmus MC Cancer Institute, UMC Rotterdam}

\date{\today}

\begin{document}
\maketitle

\begin{abstract}

Detection of minimal residual disease (MRD) in cancer patients after surgery can provide an early marker for disease recurrence and guide subsequent treatment decisions. Accurate and sensitive estimation of tumour burden after cancer surgery may be obtained through liquid biopsies, measuring circulating tumour DNA (ctDNA) using, for example, mutation-based Variant Allele Frequency (VAF) values. However, to be applicable to all patients this either requires tumour-informed, patient-specific mutation panels or sensitive, tumour-agnostic genome-wide measurements. We propose a solution that accounts for patient-specific characteristics in genome-wide screens. For that, we introduce a bivariate deconvolution model to estimate tumour proportion from circulating cell-free DNA (cfDNA) methylation profiles of patients before and after surgery. The observations are modelled as a convolution of two bivariate latent variables, corresponding to tumour and background signals, mixed by the tumour proportion at each measurement. This bivariate approach links pre- and post-surgery measurements improving estimation of the tumour proportion after surgery, when the tumour signal is potentially very weak, or absent. We approximate likelihood of the convolution through a discretisation of the bivariate density for each latent variable into a two-dimensional grid for each pair of observations which allows for fast maximum likelihood estimation. We evaluate the predictive performance of the estimated post-surgery tumour proportions based on cfDNA methylation against available mutation-based VAF values in one-year recurrence-free survival.

\end{abstract}


\section{Introduction}

After cancer surgery with curative intent, undetected cancer cells may
remain in patients which could eventually grow and lead to recurrence, months or even years after the procedure.
Identifying patients' risk of recurrence due to residual disease directly following surgery is crucial to assess the need for post-surgical treatment for every individual patient, as early detection and subsequent adjuvant treatment have better outcomes \parencite{timingACT}. 
Liquid biopsies offer a fast and minimally invasive strategy for evaluating patients during follow up. These can detect molecular biomarkers linked to the presence of cancerous cells, thereby enabling rapid screening for the presence of residual disease. Improvements in measuring techniques have increased the fidelity of tests on liquid biopsies, making less invasive metrics more reliable for clinical practice.  

Mutational profiling of the tumour tissue can be used to design highly accurate personalised assays for circulating tumour DNA (ctDNA) detection in liquid biopsies of that patient \parencite{Nakamura2024-bd, Black2025-cu}. 
Hence, notwithstanding promising results, development of these patient-specific mutation panels is complicated, expensive and requires the availability of tumour tissue. 

Tumour-agnostic approaches are available investigating for example copy number variations and cfDNA fragmentation features. However, the former require enough tumour signal to be present in order to be detected \parencite{ichorcna}, while recent evidence suggest that cfDNA fragmentation features might not be tumour-specific \parencite{CurtisFrag}. On the other hand, circulating cell-free DNA (cfDNA) methylation profiling to estimate the circulating tumour fraction is considered a promising alterative \parencite{trimethdenmark}, although standardisation is currently lacking and its clinical utility is yet to be generally validated, especially for patients whose disease burden is small \parencite{vafpredcfdna_Zhong_2024,vaftool_galant_2024}. Methylation profiles capture cell-type specific DNA methylation patterns, enabling the discrimination of circulating tumour DNA (ctDNA) from background cfDNA derived from healthy tissues. Moreover, methylation analyses have been shown to be more sensitive for detection of cancer cells in early stages \parencite{liquidmeth_kown_2023, liquidbiopreview_ma_2024}.

Several deconvolution methods have been developed to separate the tumour from background signals. \cite{demix_ahn_2013} used absolute gene expression levels for this, where they assume the background component is known. \cite{cancerloc_kang_2017} deconvoluted methylation signals by using beta distributions for cancer detection as well as for predicting the tissue-of-origin of the primary tumour. More recently, deconvolution models have been used on methylation data to estimate the proportions of different cell types in circulating cfDNA \parencite{Loyfer2023-vh, Caggiano2021-hf}. Notably, the latter three methylation-based methods operate on relative methylation values, unlike our method which uses the absolute values to respect that the actual convolution takes place on that level. Furthermore, all of the aforementioned models are univariate unlike ours.

Identifying minimal residual disease (MRD) after surgical tumour removal requires very high sensitivity due to the low tumour signal. The available deconvolution methods treat the pre- and post-operative cfDNA methylation profiles as independent. Linking pre- and post-surgery measurements, however, likely aids in the estimation of the presence of residual tumour for each individual. Our main focus is on the post-surgery estimate, yet the pre-surgery one is likely richer in tumour signal. 
To better estimate the the former, we introduce a bivariate deconvolution model with two latent components, background and tumour, weighted by each patient's respective tumour fraction. The latent variables are each assumed to be bivariate log-normally distributed, linking both measurements through a feature-wise correlation parameter. This structure allows for the baseline estimates to inform the tumour proportion estimates post surgery. We use a three-stage likelihood-based estimation process for the various parameters. 

We assessed the model's performance on simulations as well as real data and found the bivariate tumour fraction estimates more accurate than their univariate counterpart in the simulations and better linked to clinical outcomes compared to VAF-based values as well as the univariate estimates.


We use pre- and post-surgical cfDNA methylation profiles from $112$ patients with colorectal liver metastases who underwent surgery with curative intent generated using the MeD-seq assay, which generates methylated fragment counts per CpG site \parencite{medseqteo}. 
Blood samples were taken right before surgery and $3$ weeks after. Patients did not receive any peri-operative chemotherapy. We used $31$ cfDNA samples from healthy controls to $30$ samples from tumour tissue biopsies to preselect differentially methylated samples. 
Last, clinical outcome data on $1$ year recurrence-free survival (RFS) as well as overall survival (OS) is available for all patients.


\section{Model Specification}\label{sec:methods}

Suppose we observe methylated read counts of $p$ molecular features for $N$ individuals, measured immediately \textit{before} and \textit{after} surgical intervention. Let
\[
\mathbf{Y} = \left\{ (Y_{ij}^{0}, Y_{ij}^{1}) : i = 1, \ldots, N; \, j = 1, \ldots, p \right\},
\]
where $Y_{ij}^{0}$ and $Y_{ij}^{1}$ denote the expression levels of feature $j$ for individual $i$, measured pre- and post-surgery, respectively. Each observed expression level is modelled as a convolution of two independent latent components: a tumour-specific component and a background component, weighted by individual specific tumour fractions. Specifically, for each individual $i = 1, \ldots, N$, we obtain
\begin{equation}\label{eq:deconv}
\begin{aligned}
Y_{ij}^{0} &= \pi_i^{0} \, T_{ij}^{0} + (1 - \pi_i^{0}) \, B_{ij}^{0}, \\
Y_{ij}^{1} &= \pi_i^{1} \, T_{ij}^{1} + (1 - \pi_i^{1}) \, B_{ij}^{1},
\end{aligned}
\end{equation}
where $T_{ij}^{0}$ and $T_{ij}^{1}$ represent the unknown tumour-specific expression levels before and after surgery, and $B_{ij}^{0}$ and $B_{ij}^{1}$ denote the corresponding background expression levels. The parameters $\pi_i^{0}$ and $\pi_i^{1}$ represent the proportion of tumour tissue in the sample before and after surgery, respectively.

To reduce model complexity, we adopt a pseudo-likelihood approach assuming independence across the $p$ molecular features. 
While this simplification might introduce bias when strong feature correlations exist, prior work suggests that the bias is often negligible when the latent components are well-separated. Thus, the pseudo-likelihood yields consistent and computationally efficient estimators under mild conditions, justifying its use in this context \parencite{psudolik_kangro_2025, latentindentif_allman_2009}.

This assumption of independence between features allows us to model $(T_{ij}^{0}, T_{ij}^{1})$ and $(B_{ij}^{0}, B_{ij}^{1})$ for each molecular feature $j = 1, \ldots, p$ using separate bivariate log-normal distributions. Since expression levels are strictly positive and often exhibit right-skewed distributions, the (bivariate) log-normal distribution is a plausible choice for modelling molecular feature expressions \parencite{demix_ahn_2013, baylog_andrade_2021}. 

The log-normal distribution has the property that multiplying a log-normal random variable by a positive scalar results in another log-normal variable, with its mean scaled by the same factor. This enables us to model the weighted background component $(\tilde B_{ij}^{0}, \tilde B_{ij}^{1})$, where
\[
\tilde B_{ij}^{0} = (1 - \pi_i^0) B_{ij}^{0}, \quad
\tilde B_{ij}^{1} = (1 - \pi_i^1) B_{ij}^{1},
\]
with the following bivariate log-normal distribution
\begin{equation}\label{eq:bckg}
\begin{aligned}
\begin{pmatrix}
\log(\tilde B_{ij}^{0}) \\
\log(\tilde B_{ij}^{1})
\end{pmatrix}
\sim
\mathcal{N}\left(
\begin{pmatrix}
\mu_j^{0,b} + \log(1 - \pi_i^0) \\
\mu_j^{1,b} + \log(1 - \pi_i^1)
\end{pmatrix},\right. \\
\left.
\begin{pmatrix}
\tau_j^{0,b} & \tau_j^{01,b} \\
 \tau_j^{01,b}
 & \tau_j^{1,b}
\end{pmatrix}
\right),
\end{aligned}
\end{equation}
\noindent where $\mu_j^{0,b}$ and $\mu_j^{1,b}$ denote the mean log-expression levels of feature $j$ in background tissue before and after surgery, respectively. The parameters $\tau_j^{0,b}$ and $\tau_j^{1,b}$ represent the marginal variances and covariance $\tau_j^{01,b}$, with corresponding correlation $\rho_j^b = \tau_j^{01,b}/\sqrt{\tau_j^{0,b}}\sqrt{\tau_j^{1,b}}$. Then, the correlation parameter $\rho_j^b$ captures the feature-wise correlation in background expression levels between measurements. 

Similar to equation \eqref{eq:bckg}, the weighted tumour component follows a bivariate log-normal distribution with means $\mu_j^{0,t} + \log(\pi_i^0)$ and $\mu_j^{1,t} + \log(\pi_i^1)$ and marginal variances $\tau_j^{0,t}$, $\tau_j^{1,t}$, and covariance  $\tau_j^{01,t}$. Last, the feature-wise correlation parameter is given by parameter $\rho_j^t$.


\subsection{Maximum likelihood estimation}
To estimate the parameters if interest for all individuals, $\pi_i^1$, together with all the nuisance parameters, we maximise the log-pseudo-likelihood $\ell({\Pi}, {\Theta};Y) = \sum_{i=1}^n\sum_{j=1}^p \ell(\boldsymbol{\pi}_i, \boldsymbol{\theta}_j;\by_{ij})$. Here, $\quad$ $\ell(\boldsymbol{\pi}_i, \boldsymbol{\theta}_j;\by_{ij}) = \log \mathcal{L}(\boldsymbol{\pi}_i, \boldsymbol{\theta}_j;\by_{ij})$. Let $\boldsymbol{\theta}_j=\{\boldsymbol{\theta}_j^b, \boldsymbol{\theta}_j^t\}$ be the distribution parameters described above for both components, with $\boldsymbol{\theta}_j^b = \{\mu_j^{0,b}, \mu_j^{1,b}, \tau_j^{0,b}, \tau_j^{1,b}, \rho_j^{b}\}$ and equivalent for $\boldsymbol{\theta}_j^t$. Following \eqref{eq:deconv}, $\mathcal{L}(\boldsymbol{\pi}_i, \boldsymbol{\theta}_j;\by_{ij})$ is a convolution of two bivariate log-normal random variables.
Such a convolution is not analytical and available approximations have limited accuracy \parencite{baylog_andrade_2021}. It could, in principle, be computed by numerical integration, but this is likely time-consuming given that it needs to be repeated for all $j=1,\ldots, p$, all $i = 1, \ldots, n$ and all proposals of the parameters $(\Pi, \Theta)$ when optimizing
$\ell({\Pi}, {\Theta};Y)$. Therefore, we resort to a discrete approximation of the convolution.

\subsubsection{Discrete approximation of convolution}
Below we develop a discrete approximation of $\ell(\boldsymbol{\pi}_i, \boldsymbol{\theta}_j;\by_{ij})$ that allows very fast evaluation for any proposal (by the optimization algorithm) of $(\boldsymbol{\pi}_i, \boldsymbol{\theta}_j)$. For this we discretise the bivariate density for each pair $(Y_{ij}^0, Y_{ij}^1)$ with realizations $(y_{ij}^0, y_{ij}^1)$ onto a two-dimensional grid defined by all combinations of $m$ equally spaced intervals ranging from $0$ to $y^0 = y_{ij}^0$ and from $0$ to $y^1 =  y_{ij}^1$. Denote these intervals by $(C^0_r)_{r=1}^m$ and $(C^1_s)_{s=1}^m$, with
$C^0_r = [l^0_r, u^0_r] = [(r-1)y^{0}/m, ry^{0}/m ]$ and $C^1_s = [l^1_s, u^1_s] = [(s-1)y^{1}/m, sy^{1}/m ]$. Here, the last intervals, $C_m^0$ and $C_m^1$ correspond to realizations $y^0$ and $y^1$,  respectively. 
We then create the same grids for both pairs $(\tilde{T}^1, \tilde{T}^2)$ and  $(\tilde{B}^1, \tilde{B}^2)$, with $\tilde{T}^k = \tilde{T}_{ij}^k, \tilde{B}^k = \tilde{B}_{ij}^k.$
These grids are depicted in Figure \ref{fig:grid}.
Then, we have for $\boldsymbol{\pi} = \boldsymbol{\pi}_i$ and $\boldsymbol{\theta} = \boldsymbol{\theta}_j$:

\begin{figure}
    \centering
    \includegraphics[width=0.95\linewidth]{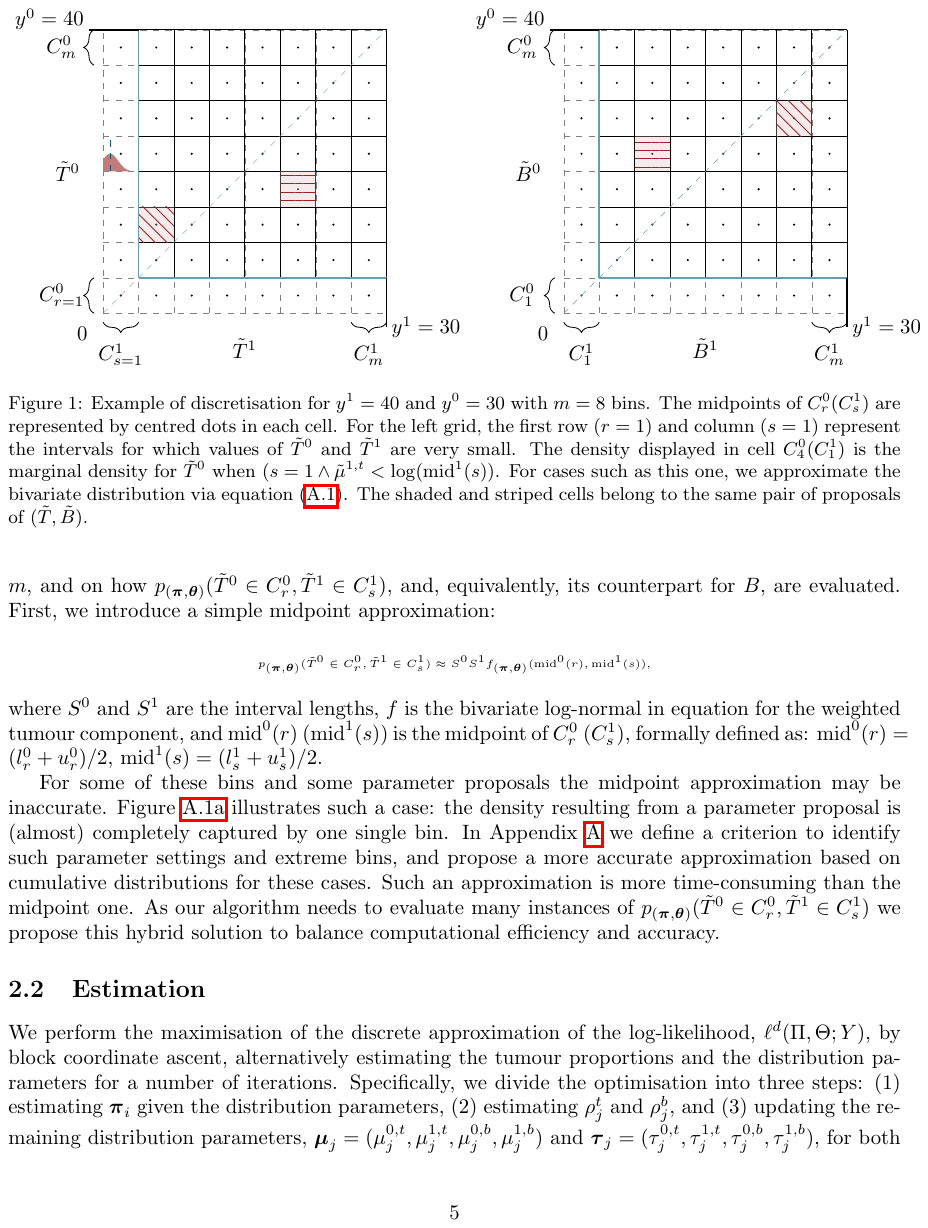}
    \caption{Example of discretisation for $y^1 = 40$ and $y^0=30$ with $m=8$ bins.  The midpoints of $C_r^0(C_s^1)$ are represented by centred dots in each cell. For the left grid, the first row ($r=1$) and column ($s=1$) represent the intervals for which values of $\tilde{{T}}^0$ and $\tilde{{T}}^1$ are very small. The density displayed in cell $C_4^0(C_1^1)$ is the marginal density for $\tilde{{T}}^0$ and illustrates a case for which the midpoint approximation is likely inaccurate.. The shaded and striped cells belong to the same pair of proposals of $({\tilde{T}}, {\tilde{B}})$.}
    \label{fig:grid}
\end{figure}

\begin{equation} \label{eq:cdfEval}
\begin{split}
p_{(\boldsymbol{\pi}, \boldsymbol{\theta})} &(Y^{0} \in C^0_m, Y^{1} \in C^1_m)\\ &\approx \sum_{r=1}^{m} \sum_{s=1}^{m} \Bigl(p_{(\boldsymbol{\pi}, \boldsymbol{\theta})}(\tilde{T}^{0} \in C^0_r, \tilde{T}^{1} \in C^1_s) \\ & \times   p_{(\boldsymbol{\pi}, \boldsymbol{\theta})}(\tilde{B}^{0} \in C^0_{m+1-r}, \tilde{B}^{1} \in C^1_{m+1-s})\Bigr),
\end{split}
\end{equation}
which uses the independence of $\tilde{T}$ and $\tilde{B}$.

Then, applying equation \eqref{eq:cdfEval} to all $(i,j)_{i=1,j=1}^{n,p}$ defines a discrete log-likelihood $\ell^d({\Pi}, {\Theta};Y)$ that approximates
$\ell({\Pi}, {\Theta};Y)$. Here, the accuracy of the approximation depends on the number of bins, $m$, and on how
$p_{(\boldsymbol{\pi}, \boldsymbol{\theta})}(\tilde{T}^{0} \in C^0_r, \tilde{T}^{1} \in C^1_s)$, and, equivalently, its counterpart for $B$, are evaluated. First, we introduce a simple midpoint approximation: 
\begin{equation} \label{eq:midpoint}
\begin{split}\tiny
    p_{(\boldsymbol{\pi}, \boldsymbol{\theta})}(\tilde{T}^{0} \in C^0_r, \tilde{T}^{1} \in C^1_s) \approx 
    S^0 S^1 f_{(\boldsymbol{\pi}, \boldsymbol{\theta})}(\text{mid}^0(r),\text{mid}^1(s)), \nonumber
\end{split}
\end{equation}
where $S^0$ and $S^1$ are the interval lengths, $f$ is the bivariate log-normal in equation for the weighted tumour component, and
$\text{mid}^0(r)$ ($\text{mid}^1(s)$) is the midpoint of $C^0_r$ ($C^1_s$), formally defined as:
$\text{mid}^0(r) = (l^0_r + u^0_r)/2$, $\text{mid}^1(s) = (l^1_s + u^1_s)/2$.

For some of these bins and some parameter proposals the midpoint approximation may be inaccurate. Figure \ref{fig:cuma} illustrates such a case: the density resulting from a parameter proposal is (almost) completely captured by one single bin. In Appendix \ref{app:cumulative} we define a criterion to identify such parameter settings and extreme bins, and propose a more accurate approximation based on cumulative distributions for these cases. Such an approximation is more time-consuming than the midpoint one. As our algorithm needs to evaluate many instances of $p_{(\bpi, \btheta)}(\tilde{T}^0 \in C_r^0, \tilde{T}^1 \in C_s^1)$ we propose this hybrid solution to balance computational efficiency and accuracy.


\subsection{Estimation}
We perform the maximisation of the discrete approximation of the log-likelihood, $\ell^d({\Pi}, {\Theta};Y)$, by block coordinate ascent, alternatively estimating the tumour proportions and the distribution parameters for a number of iterations. Specifically, we divide the optimisation into three steps: (1) estimating $\boldsymbol{\pi}_i$ given the distribution parameters, (2) estimating $\rho_j^t$ and $\rho_j^b$, and (3) updating the remaining distribution parameters, $\bmu_j=(\mu_j^{0,t}, \mu_j^{1,t}, \mu_j^{0,b}, \mu_j^{1,b})$ and $\btau_j=(\tau_j^{0,t}, \tau_j^{1,t}, \tau_j^{0,b}, \tau_j^{1,b})$, for both components given the correlation coefficients and tumour proportions. For identifiability we impose the plausible constraints $\pi_i^0 > \pi_i^1$ $\forall i$. The initial values for $\boldsymbol{\mu}_j$ and $\boldsymbol{\tau}_j$ are given by methylation profiles generated from tumour tissues and healthy cfDNA respectively, while $\rho_j^b$ and $\rho_j^t$ may be initialised randomly.

\section{Results}
\subsection{Simulations} \label{sec:sim_deconv}


We simulated our data for $100$ patients and $600$ features following equation \eqref{eq:deconv}. For each latent variable, we draw the distribution parameters from densities based on real tumour and background methylation profiles from tissue biopsies and healthy cfDNA, respectively. We kept $\mu_j^t$ equal for both measurements but allow $\mu_j^{l_0}$ to differ from $\mu_j^{l_1}$ and $\tau^{l_0}$ from $\tau^{l_1}$ to make the simulated data more realistic. This differences may appear since interventions such as surgery are associated with changes in DNA methylation, particularly those linked to immune regulation \parencite{chnagesgeneexprsurgery_sadahiro_2020}. We used available data on the expected ctDNA levels to simulate $\bpi^{0}$ and $\bpi^{1}$ \parencite{miracleOG}, then setting half of the latter to zero (Figure \ref{fig:sim_pi}). Last, the correlation values were drawn from uniform distributions $\rho_j^t \sim \mathcal{U}(0.9, 1)$ and $\rho_j^l \sim \mathcal{U}(0.5, 1)$. The rationale behind the choices of lower bounds is that tumour methylation profiles are expected to remain relatively stable with respect to their baseline values, whereas the background signal may become more heterogeneous as a result of the surgical intervention.

We estimated the tumour distribution parameters and the proportions at both measurements using the scheme described in section \ref{sec:methods} over $10$ iterations. 
To evaluate the benefits of including the correlation parameter, we estimated two additional specifications, one with assumed shared correlation estimates for the latent variables, $\rho_j^t = \rho_j^l$ and another one where $\rho_j^t = \rho_j^l = 0$ $\forall j$. The association between estimated tumour proportions, $(\hat{\pi}_i^0, \hat{\pi}_i^1)$, and the true proportions are collected in Table \ref{tab:sim_cor} and displayed in Figure \ref{fig:sim_cor} in the Appendix. For all three specifications, we initialised $\boldsymbol{\Theta}$ using the true parameter values from the tissue biopsy, and correlations of $0.8$ for the ones using the correlation parameter. As expected, the association for $\hat{\pi}_i^0$ is similar for all specifications with the one including separate correlation coefficients having a lower Root Mean Squared Error (RMSE), compared to the other two. However, for the post-surgery tumour proportions, $\pi_i^1$, for which the tumour signal is likely weaker or not present, the benefits from incorporation $\rho_j^b$ and $\rho_j^t$ are clearer, with both associations with the true proportions as well as RMSE showing improvements compared to the specification assuming no correlation.

\begin{table}[!h]
    \centering 
\begin{tabular}{lccc|ccc}
\hline\hline
\multicolumn{1}{c}{} &
\multicolumn{3}{c}{$\hat{\bpi}^0$} &
\multicolumn{3}{c}{$\hat{\bpi}^1$} \\
\cline{2-7}

 &  \textbf{(A)}& \textbf{(B)}& \textbf{(C)}&  \textbf{(A)}& \textbf{(B)}& \textbf{(C)} \\ 
Assoc. & 0.991 & 0.988 &  0.990 & 0.951 & 0.952 & 0.885 \\
RMSE   & 0.020 & 0.025 &  0.024 & 0.011 & 0.012 & 0.017  \\

\hline\hline
\end{tabular} \vspace{1em}
\captionof{table}{Association (Assoc.) of true and estimated tumour proportions for each specification. Column \textbf{(A)} collects the results for$\hat{\rho}_j^t \neq \hat{\rho}_j^l $, column \textbf{(B)} for $\hat{\rho}_j^t = \hat{\rho}_j^l $, and column \textbf{(C)} for the univariate case, $\hat{\rho}_j^t = \hat{\rho}_j^l =0$.}
    \label{tab:sim_cor}
\end{table}

We performed two different robustness checks to validate our model. First, we obtained our estimates given different initial values to analyse their effect on the estimated tumour proportions and evaluate convergence of distribution parameters estimates to their true values. Even though initial values for $\boldsymbol{\mu}_j$ and $\boldsymbol{\tau}_j$  are usually available, factors such as measurement error may influence the reported quantities, hindering their accuracy. Hence, we further assessed our model using different initialisation values, setting $\boldsymbol{\mu}_j^{b} = 1.5$ and $\boldsymbol{\mu}_j^{t} = 3$, for all features in both measurements. The former is approximately the average of the background feature means and the latter is slightly below the tumour average of all means. Second, we simulated data with uncorrelated measurements to determine if including said correlation coefficients hamper the overall performance of the model. Overall, including correlation estimates showed better adherence to the true tumour proportions when the initial values did not match the true distribution values and did not disrupt the performance under the uncorrelated measurement scenario. Results on the RMSE and the association between the estimated and the true tumour proportions post-surgery for these checks are collected in Figures \ref{fig:sim_cor_mut3} and \ref{fig:sim_cor_0rho_mut3_truerho0} in Appendix \ref{app:sim_cor}. Altogether, the bivariate specification performed better under constant initial values and did not show loss of power if the measured data was uncorrelated.

Last, we validate the convergence of the estimated parameters under the aforementioned initial values over $100$ iterations (Figure \ref{fig:convergence} in Appendix \ref{app:convergence}). 
Altogether, the bivariate deconvolution approach outperformed its univariate counterpart particularly estimating the post-surgery tumour proportions, without losing accuracy when the latent variables are not correlated between measurements.

\subsection{Real data} \label{sec:data_deconv}

\noindent We then applied our model on a real dataset of $112$ patients with colorectal liver metastases (CRLM) undergoing liver resection \parencite{miracleOG} profiled using the \textit{MeD-seq} assay. We used TPM-normalised methylation counts from $1069$ CpG islands that we had previously identified as differentially methylated between CRLM tissue and healthy cfDNA. 
The samples were collected directly before surgery and $3$ weeks after the intervention, and none of the patients received peri-operative systemic treatment. We preselected the $832$ regions that were significantly hypermethylated in the tumours with log fold change larger than $1$ a common cut-off strategy \parencite{Willems2016-el, Da_Mota2026-he}. Additionally, we used available MeD-seq data from $31$ healthy cfDNA and $30$ tumour tissues \parencite{stavrosSaskia} to initialise $\boldsymbol{\mu}_j$ and $\boldsymbol{\tau}_j$. We used one-year RFS outcomes to assess our tumour fraction estimates' link to patient relapse against the mutation-based tumour fractions. 

\begin{figure}[!h]
    \centering
    \includegraphics[width=0.5\linewidth]{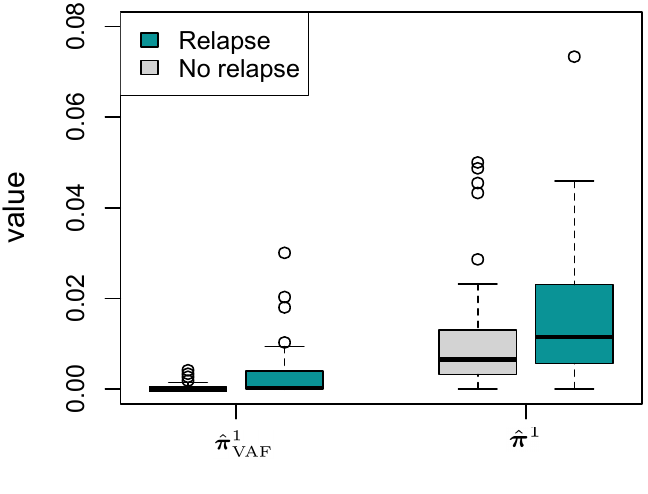}
    \caption{$\hat{\bpi}^1_{\text{VAF}}$ and $\hat{\bpi}^1$.}
    \label{fig:boxplot}
\end{figure}

The distribution of the post-surgery VAF-based values, denoted by $\hat{\bpi}^1_{\text{VAF}}$, and estimated tumour proportions are displayed in Figure \ref{fig:boxplot}. The former were available for $71$ of the $112$ patients while we estimated the latter for all individuals. Of the $112$ patients, half experienced a recurrence event within one year. Among the patients with available $\hat{\bpi}^1_{\text{VAF}}$, $33$ did not relapse in the year following the intervention, versus $38$ who did. For the sake of comparison, we use data from patients for whom both VAF and methylation data are available. 

We analysed the link between $\hat{\bpi}^1_{\text{VAF}}$ and estimated tumour fractions and  one-year RFS, which is a binary outcome, as censoring did not occur within one year. 
Figure \ref{fig:boxplot} shows that $\hat{\bpi}^1_{\text{VAF}}$ were largely concentrated near zero, regardless of the outcome, while the post-surgery tumour proportion estimates displayed more contrast between groups. 

We performed a logistic regression for both predictors to evaluate their link with the one-year RFS outcomes. Due to the zero-inflated values of the predictors, particularly $\hat{\bpi}^1_{\text{VAF}}$, we dichotomised them at the median, categorising the values as either above or below this threshold. For consistency, we dichotomise both $\hat{\bpi}^1_{\text{VAF}}$ as well as our estimates. 
For the logistic regression, the VAF predictor was not significantly associated with the outcome. Conversely, the estimated tumour proportions showed a significant link, with patients with above median values having three time higher odds of experiencing a relapse within the first year after surgery, compared to their counterparts. Although significant, the small sample size limits precision assertions on the magnitude of the association between the tumour fraction estimated and the clinical outcome ($OR=3.429$, 95\%CI $\left[1.314, 9.404\right]$). The Wilcoxon and Fisher tests corroborated these findings for the respective associations with the outcome. ROC analysis further supported the improved predictive ability of our model's tumour proportion estimate to separate between the relapsed and the recurrence-free patients (see Table \ref{tab:sumres}). 
We extended this analysis to all patients to comparing the estimated tumour fractions from the bivariate deconvolution to their univariate counterparts. Supplementary Table \ref{tab:sumres_app} shows
the latter was not significantly linked to RFS outcomes while the bivariate estimate was ($OR=2.388$, 95\%CI $\left[1.128, 5.164\right]$). These results were confirmed by the Wilcoxon and Fisher tests. 
In line with the results from the simulations, we observe the bivariate model estimate as a reliable predictor for relapse. Furthermore, these estimates show a clearer distinction between groups, particularly when compared to the univariate estimates.



\begin{table}[!h]
    \centering 
\begin{tabular}{lcccc}
\hline\hline
 &  $\hat{\bpi}^1_{\text{VAF}}$ && $\hat{\bpi}^1$ \\ 
\hline
Wilcoxon ($p$-value) & 0.034 &&  0.005 \\
Fisher ($p$-value)   & 0.150 &&  0.017  \\
Logistic ($\beta$)   & 0.77  &&  1.23*  \\
AUC                  & 0.618 &&  0.693 \\
\hline\hline
\end{tabular} 
\captionof{table}{Comparing $\hat{\bpi}^1_{\text{VAF}}$ to $\hat{\bpi}^1$.}
    \label{tab:sumres}
\end{table}



\begin{figure}[!h]
    \centering
    \begin{subfigure}[t]{0.45\textwidth}
        \centering
        \includegraphics[height=1.5in]{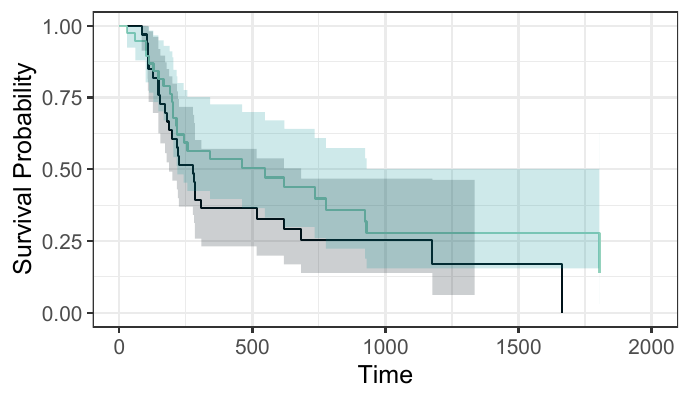}
        \caption{RFS curves for VAF values.}
    \end{subfigure}%
    ~ 
    \begin{subfigure}[t]{0.45\textwidth}
        \centering
        \includegraphics[height=1.5in]{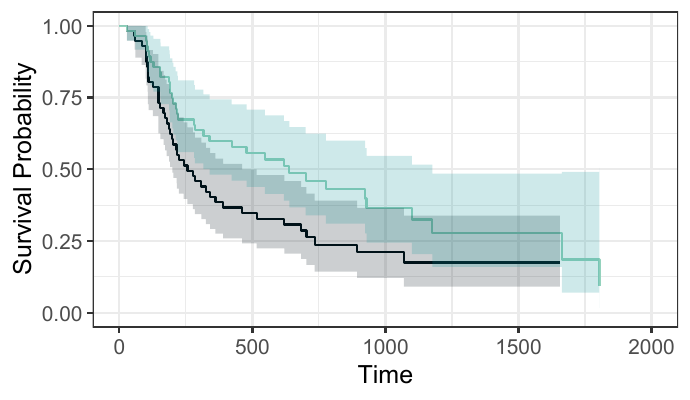}
        \caption{RFS curves for $\boldsymbol{\hat{\pi}}^1$.}
    \end{subfigure}
    \begin{subfigure}[t]{0.45\textwidth}
        \centering
        \includegraphics[height=1.5in]{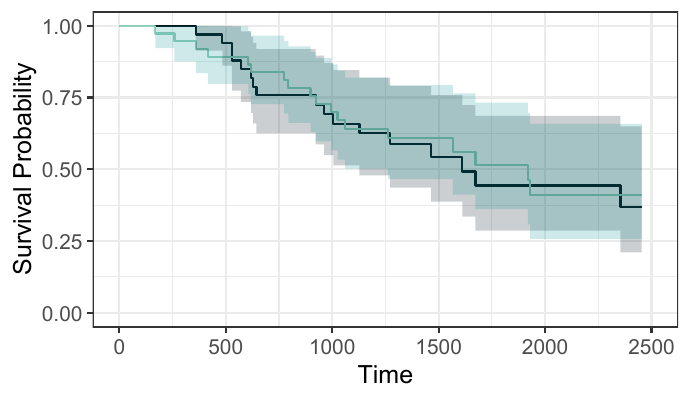}
        \caption{OS curves for VAF values.}
    \end{subfigure}
    ~ 
    \begin{subfigure}[t]{0.45\textwidth}
        \centering
        \includegraphics[height=1.5in]{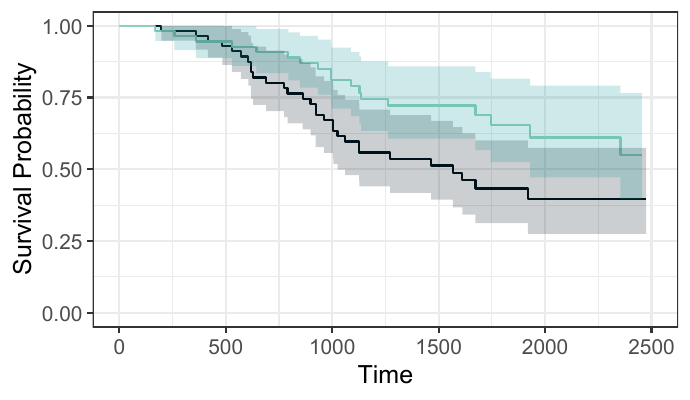}
        \caption{OS curves for $\boldsymbol{\hat{\pi}}^1$}
    \end{subfigure}
    ~ 
    \begin{subfigure}[t]{0.45\textwidth}
        \centering
        \includegraphics[height=0.5in]{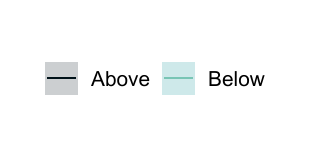}
        \caption{OS curves for $\boldsymbol{\hat{\pi}}^1$}
    \end{subfigure}
    \caption{Kaplan-Meier curves for OS, in days.}
    \label{fig:kmc_os}
\end{figure}

Last, we calculated the Kaplan-Meier curves for the RFS and OS for $\hat{\bpi}^1_{\text{VAF}}$ and $\hat{\bpi}^1$ are shown in Figure \ref{fig:kmc_os}. Once more, our estimate is able to separate the trajectories for each group trough time in both cases better than $\hat{\bpi}^1_{\text{VAF}}$. Altogether, these results show the improvements in accuracy of our tumour fraction estimates' association to clinical outcomes. 

\section{Discussion} \label{sec:conclusion}

Monitoring cancer patients after surgery is required to assess risk of recurrence and determine their treatment needs, as timely detection in order to provide adequate treatment is crucial. 
However, early detection of MRD remains difficult, as tumour signal is hard to detect under quick diagnostic tools or requires procedures too invasive for continuous screenings. 
Moreover, sensitivity is essential in the MRD setting, where ctDNA load is very low, due to the main bulk of the tumour having been removed and surgical trauma further diluting the signal. 

We introduced a bivariate deconvolution model to estimate the residual disease from post-surgical cfDNA methylation profiles. This approach links the pre- and post-surgery measurements via feature-wise correlation of the two latent components. The simulation study and the real data analysis confirmed the benefits from linking both measurements with a feature-wise correlation parameter when compared to the results from a univariate deconvolution as well as available $\hat{\bpi}^1_{\text{VAF}}$ for the real data. Moreover, when the simulated data was uncorrelated, the bivariate specification did not result in meaningful decrease in accuracy. This correlation structure is a reasonable one, as treatment is absent between measurements in our setting. As this may not be the case in many cancer monitoring scenarios, the model can account for a global treatment effect by allowing a possible difference in means per methylation site, for each measurement. 
This may require larger sample sizes or additional parameter constraints to yield reliable estimates. 


In this paper we focused on methylation data obtained using the MeD-seq assay, which only generates \textit{methylated} read counts, as opposed to whole genome bisulphite sequencing (WGBS), where the number of \textit{unmethylated} fragments is also known. In order to adapt this method to such data, one could potentially extend this bivariate approach into a four-variate model. 

Finally, incorporating identification of cell-type heterogeneity could further detail tumour and background compositions and improve current understanding of how treatment may affect the tumour and background components at different levels.

\printbibliography

\appendix

    \counterwithin{figure}{section}
    \counterwithin{table}{section}
    \counterwithin{equation}{section}

\clearpage

\section{Estimation of edge cells} \label{app:cumulative}
For tumour fraction $\pi^1$ that is very small (which is not unlikely for some samples) and $y^1$ is fairly large. Then, much of the probability mass of $\tilde{T}^1$ is likely captured by bins for which $s=1$, which renders the midpoint approximation to be rather inferior. Denote the mean of $\log(\tilde{T}^k)$ by $\tilde{\mu}^{k,t} = \mu^{k,t} + \log(\pi^k)$. Formally, we identify such bins by the following criterion:
\begin{equation*}\tiny
    (r=1 \land \tilde{\mu}^{0,t} < \log(\text{mid}^1(r))) \lor (s=1 \land \tilde{\mu}^{1,t} < \log(\text{mid}^1(s))).
\end{equation*}

\begin{figure}[!htb]
    \centering
        \begin{subfigure}[b]{0.48\linewidth}
        \includegraphics[height=2.2in]{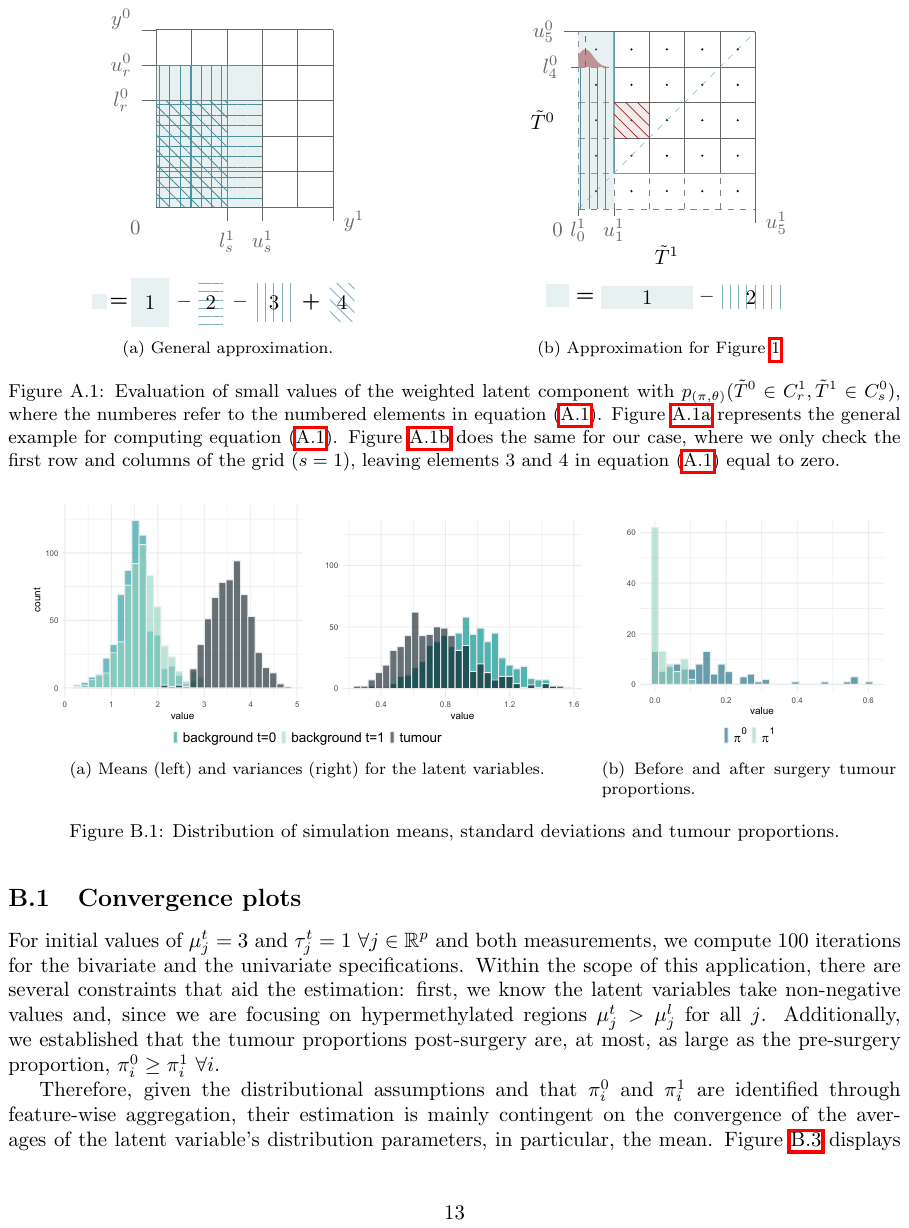}
        \caption{General Approximation}
        \label{fig:cuma}
    \end{subfigure} %
\hfill
    \begin{subfigure}[b]{0.48\linewidth}    
        \includegraphics[height=2.2in]{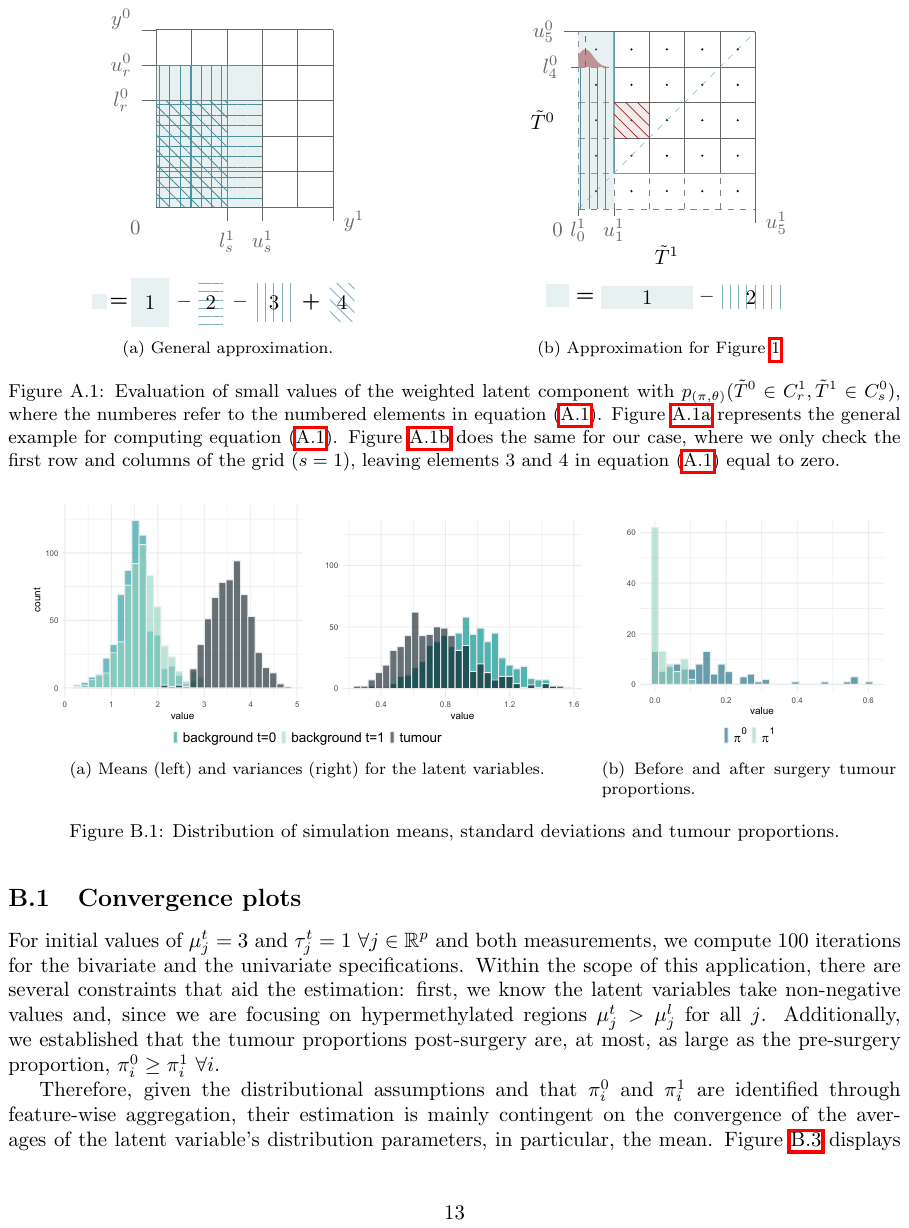}
        \caption{Approximation for Figure \ref{fig:grid}}
        \label{fig:cumb}    
    \end{subfigure} 
    \caption{Evaluation of small values of the weighted latent component with $p_{({\pi}, {\theta})}(\tilde{T}^{0} \in C^1_r, \tilde{T}^{1} \in C^0_s)$, where the numbers refer to the numbered elements in equation \eqref{eq:cumulative}. Figure \ref{fig:cuma} represents the general example for computing equation \eqref{eq:cumulative}. Figure \ref{fig:cumb} does the same for our case, where we only check the first row and columns of the grid ($s=1$), leaving elements $3$ and $4$ in equation \eqref{eq:cumulative} equal to zero.}
    \label{fig:gridcum}
\end{figure}

As these bins capture a large part of $\tilde{T}$'s distribution, we evaluate them by
\begin{equation}\label{eq:cumulative}
\begin{split}
p_{(\boldsymbol{\pi}, \boldsymbol{\theta})}(\tilde{T}^{0} \in C^0_r, \tilde{T}^{1} \in C^1_s)&=
\int_{l^0_r}^{u^0_r} \int_{l^1_s}^{u^1_s}  f_{(\boldsymbol{\pi}, \boldsymbol{\theta})}(t_0,t_1) dt^0dt^1\\
&= \underbrace{F_{(\boldsymbol{\pi}, \boldsymbol{\theta})}(u^0_r, u^1_s)}_{1} - \underbrace{F_{(\boldsymbol{\pi}, \boldsymbol{\theta})}(l^0_r, u^1_s)}_{2} 
- \underbrace{F_{(\boldsymbol{\pi}, \boldsymbol{\theta})}(u^0_r, l^1_s)}_{3}  +  \underbrace{F_{(\boldsymbol{\pi}, \boldsymbol{\theta})}(l^0_r, l^1_s)}_{4}.
\end{split}
\end{equation} 
instead, with $F$ denoting the bivariate log-normal CDF. Such CDFs are more tedious to compute than the PDFs required for the mid-point approximation. For our algorithm the number of evaluations of such probabilities equals $2\times m^2\times n\times p\times I,$ with $I$ the number of proposals by the optimizer. As this number tends to be very large, we only employ the CDFs for the extreme bins. For bins with $r>1$ or $s>1$ it is very unlikely that a very large part of the marginal distribution is captured by the bin, so therefore the midpoint approximation suffices for these bins. See Figure \ref{fig:gridcum} for an example. These are the bins corresponding to $r=1$ or $s=1$, which reduces the second and fourth terms of the above approximation to zero, as shown in Figure \ref{fig:cumb}. 
Figures \ref{fig:cuma} and \ref{fig:cumb} represent the evaluation of the weighted components at bins for which the midpoint approximation may be inaccurate, due to, most likely, the tumour fraction being close to zero. These figures match the evaluation described in equation \eqref{eq:cumulative} in the main text. 

\section{Additional simulations results} \label{app:sim}

\begin{figure}[!htb]
    \centering
    \begin{subfigure}[t]{0.66\linewidth}
        \centering
        \includegraphics[width=0.95\linewidth]{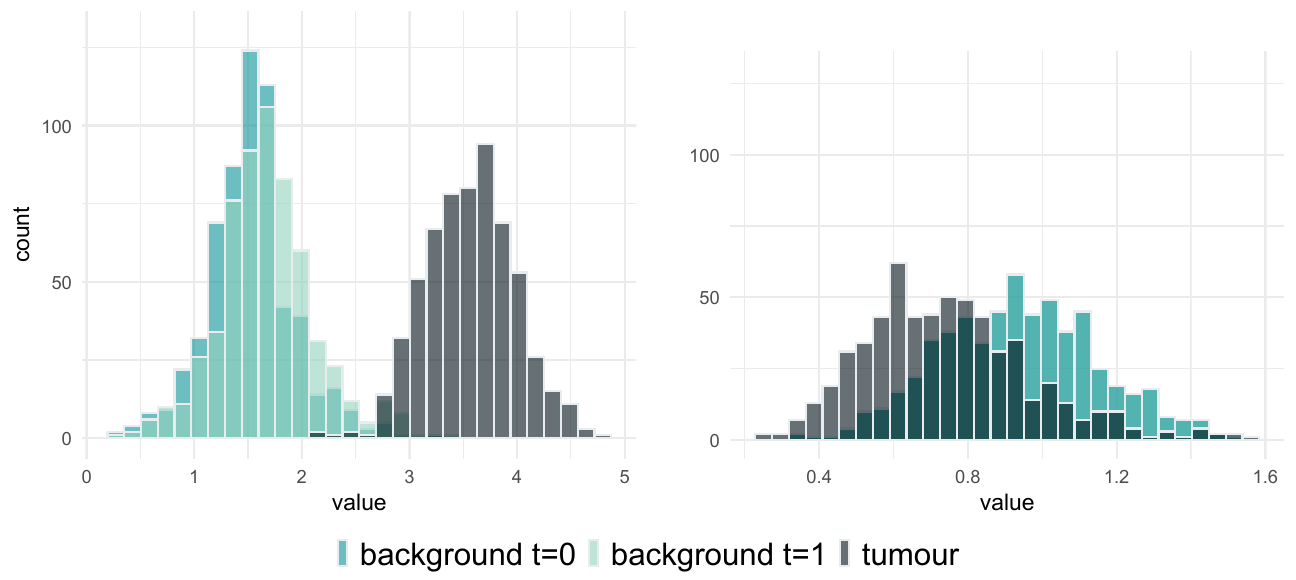}
    \caption{Means (left) and variances (right) for the latent variables.}
    \label{fig:sim_theta}
    \end{subfigure}%
    \hfill
    \begin{subfigure}[t]{0.33\linewidth}
        \centering
        \includegraphics[width=0.95\linewidth]{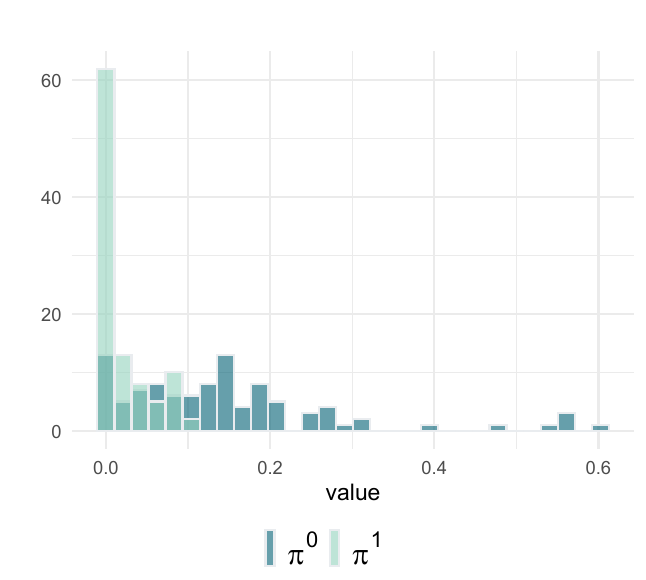}
        \caption{Before and after surgery tumour proportions.}
        \label{fig:sim_pi}
    \end{subfigure}
    \caption{Distribution of simulation means, standard deviations and tumour proportions.}
    \label{fig:sim_theta_pi}
\end{figure}

\begin{figure*}[!htb]
    \centering
    \begin{subfigure}[t]{0.3\textwidth}
        \centering
        \includegraphics[width=0.95\linewidth]{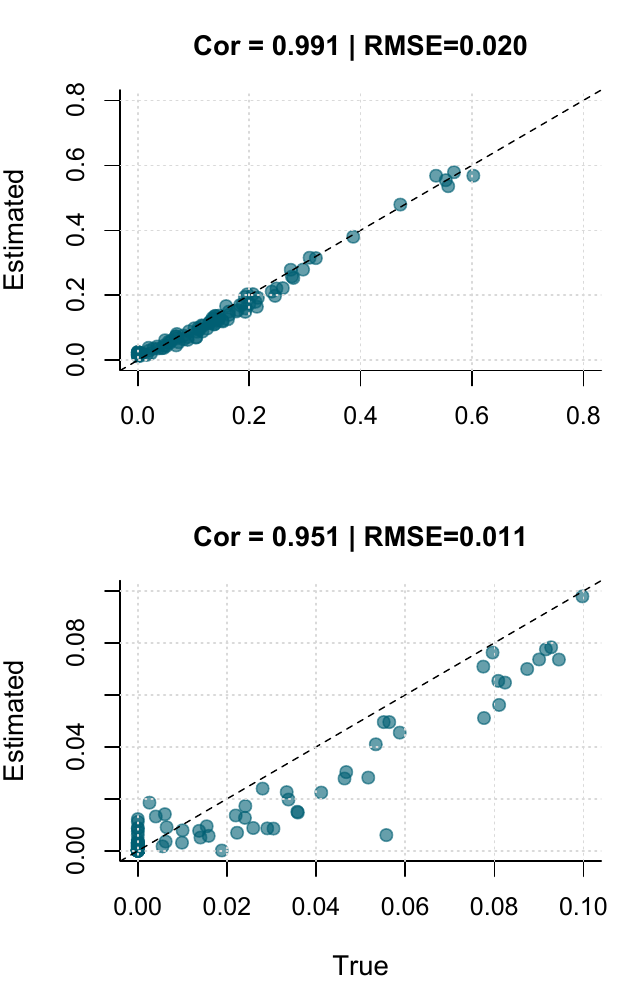}
    \caption{$\hat{\rho}_j^t \neq \hat{\rho}_j^l $}
    \label{fig:sim_cor_2rho}
    \end{subfigure}%
    \hfill
    \begin{subfigure}[t]{0.3\textwidth}
        \centering
        \includegraphics[width=0.95\linewidth]{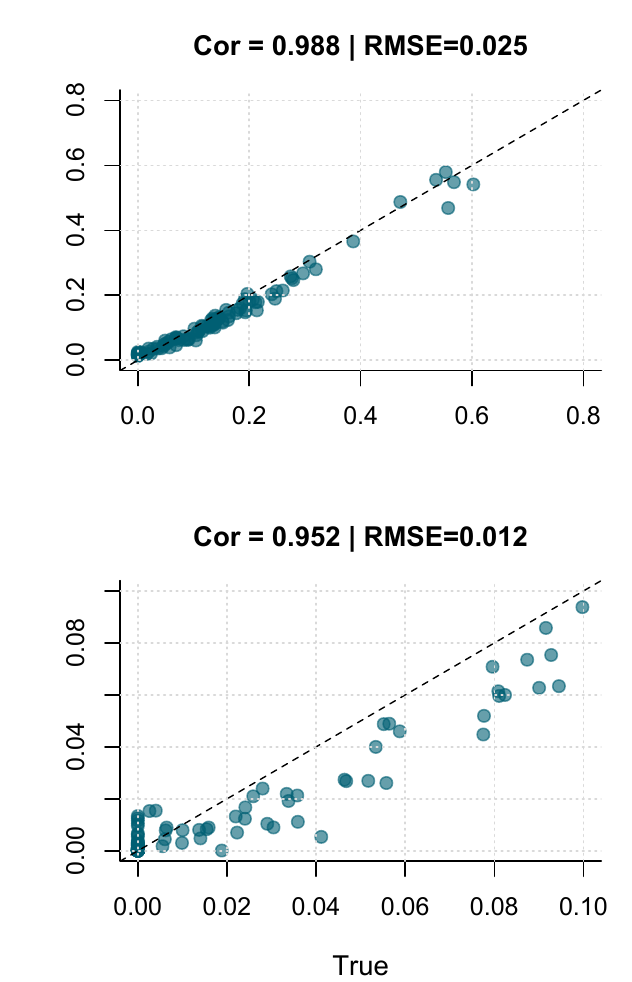}
    \caption{$\hat{\rho}_j^t =\hat{\rho}_j^l $}
    \label{fig:sim_cor_1rho}
    \end{subfigure}%
    \hfill
    \begin{subfigure}[t]{0.3\textwidth}
        \centering
        \includegraphics[width=0.95\linewidth]{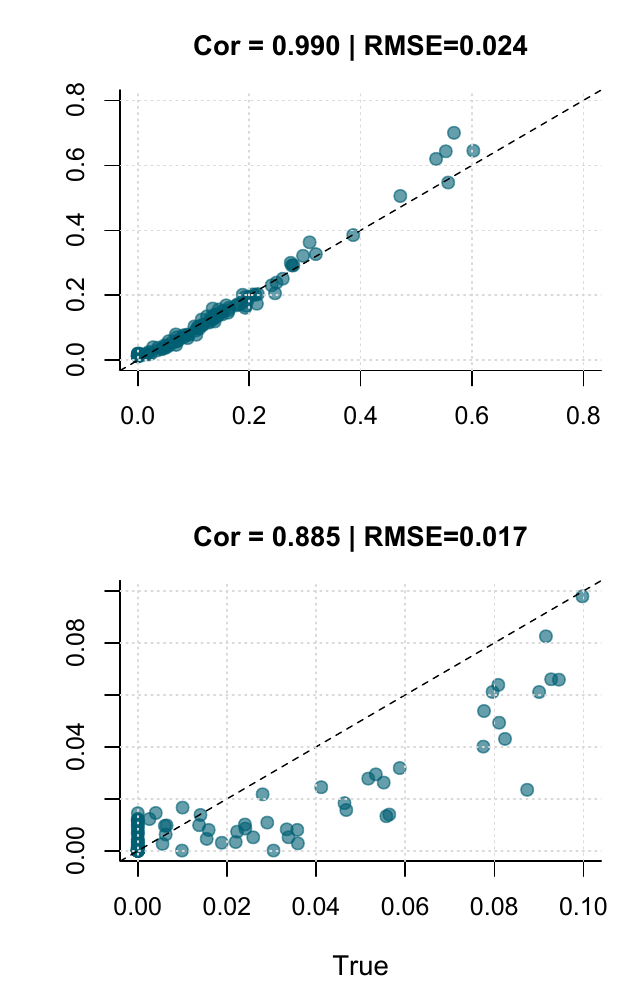}
        \caption{$\hat{\rho}_j^t =\hat{\rho}_j^l  = 0$}
        \label{fig:sim_cor_0rho}
    \end{subfigure}
    \caption{Association of true and estimated tumour proportions, $\boldsymbol{\hat{\pi}}^0$ (top) and  $\boldsymbol{\hat{\pi}}^1$  (bottom).}
    \label{fig:sim_cor}
\end{figure*}

We simulated our data in Section \ref{sec:sim_deconv} following equation \eqref{eq:sim_deconv}, where $\epsilon^0_{ij}$ and $\epsilon^1_{ij}$ represent noise terms. The latent components are modelled based on the feature means and standard deviations from tissue biopsies from each component, as seen in Figure \ref{fig:sim_theta}. The tumour proportions in Figure \ref{fig:sim_pi} are based of the pre- and post-surgery$\hat{\bpi}^1_{\text{VAF}}$. Figure \ref{fig:sim_cor} displays de scatter plot for the association and RMSE values shown in Table \ref{tab:sim_cor}.

\begin{equation}
    \begin{aligned} \label{eq:sim_deconv}
        \log y^0_{ij} &= \log\bigl(\pi^0_iT^0_{ij} + (1-\pi^0_i)B^0_{ij}\bigr) + \epsilon^0_{ij} \\
        \log y^1_{ij} &= \log\bigl(\pi^1_iT^1_{ij} + (1-\pi^1_i)B^1_{ij}\bigr) + \epsilon^1_{ij}.
    \end{aligned}
\end{equation}

\noindent Figure \ref{fig:sim_theta} displays the tumour tissue and cfDNA background means and standard deviations used to simulated the data. The background means have noise added to increase distortion that may come from surgery or other treatments. The simulated tumour fractions are in Figure \ref{fig:sim_pi}. Last, Figure \ref{fig:sim_cor} shows the association plots for the simulations described in Section \ref{sec:sim_deconv} and collected in Table \ref{tab:sumres}.

\subsection{Convergence plots} \label{app:convergence}
For initial values of $\mu_j^t=3$ and $\tau_j^t=1$ $\forall j\in \mathbb{R}^p$ and both measurements, we compute $100$ iterations for the bivariate and the univariate specifications. Within the scope of this application, there are several constraints that aid the estimation: first, we know the latent variables take non-negative values and, since we are focusing on hypermethylated regions $\mu_j^t >\mu_j^l$ for all $j$. Additionally, we established that the tumour proportions post-surgery are, at most, as large as the pre-surgery proportion, $\pi_i^0 \geq \pi_i^1$ $\forall i$.

\begin{figure}[!h]
\centering
    \begin{subfigure}[t]{0.45\textwidth}
        \centering
        \includegraphics[width=0.95\linewidth]{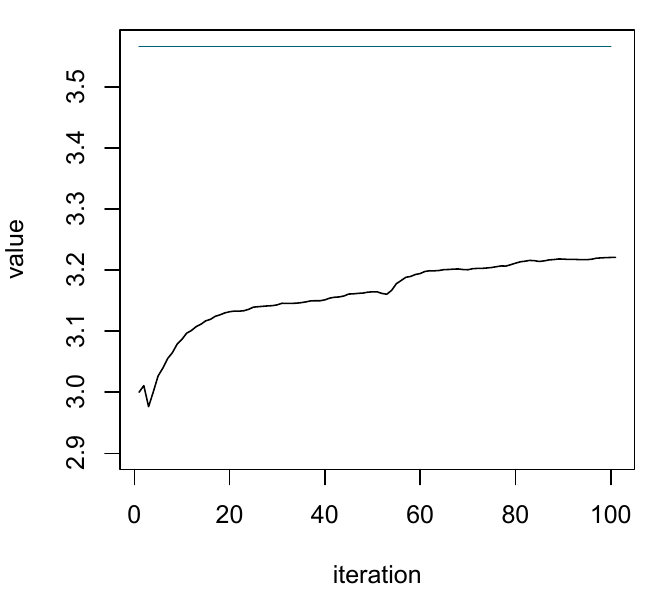}
    \caption{$\hat{\rho}_j^t \neq \hat{\rho}_j^l $}
    \label{fig:sim_cor_1rho}
    \end{subfigure}%
    \hfill
    \begin{subfigure}[t]{0.45\textwidth}
        \centering
        \includegraphics[width=0.95\linewidth]{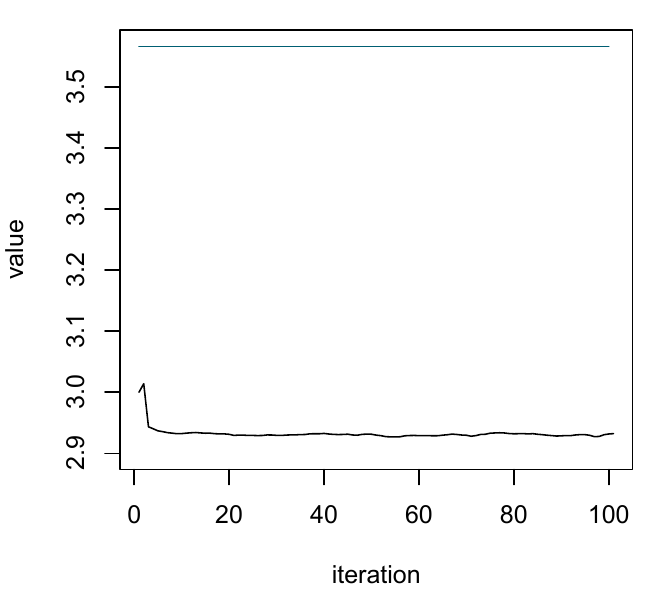}
        \caption{$\hat{\rho}_j^t = \hat{\rho}_j^l $}
        \label{fig:sim_cor_0rho}
    \end{subfigure}
    \label{fig:sim_cor_1rhosim_cor_mut3_mut15}
    \caption{Tumour means convergence plot. Average convergence of the tumour means for the bivariate and univariate specifications, respectively. The blue horizontal line represents the true average mean.}
    \label{fig:convergence}
\end{figure}

Therefore, given the distributional assumptions and that $\pi_i^0$ and $\pi_i^1$ are identified through feature-wise aggregation, their estimation is mainly contingent on the convergence of the averages of the latent variable's distribution parameters, in particular, the mean. Figure \ref{fig:convergence} displays the average convergence of the tumour means for the bivariate (left) and the univariate (right) specifications when the data are correlated. Keeping all other factors equal, the bivariate specification shows convergence on average of the tumour means while the univariate counterpart stays close to the initial values.


\subsection{Robustness checks}  \label{app:sim_cor}
\begin{figure*}[!h]
    \centering
    \begin{subfigure}[t]{0.3\textwidth}
        \centering
        \includegraphics[width=0.95\linewidth]{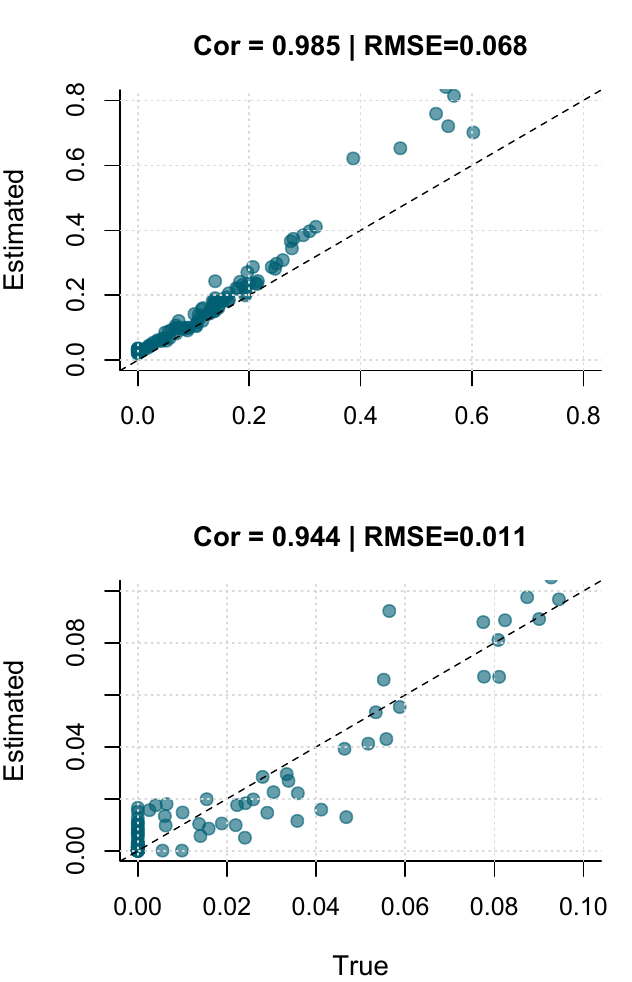}
    \caption{$\hat{\rho}_j^t \neq \hat{\rho}_j^l $}
    \label{fig:sim_cor_2rho}
    \end{subfigure}%
    \hfill
    \begin{subfigure}[t]{0.3\textwidth}
        \centering
        \includegraphics[width=0.95\linewidth]{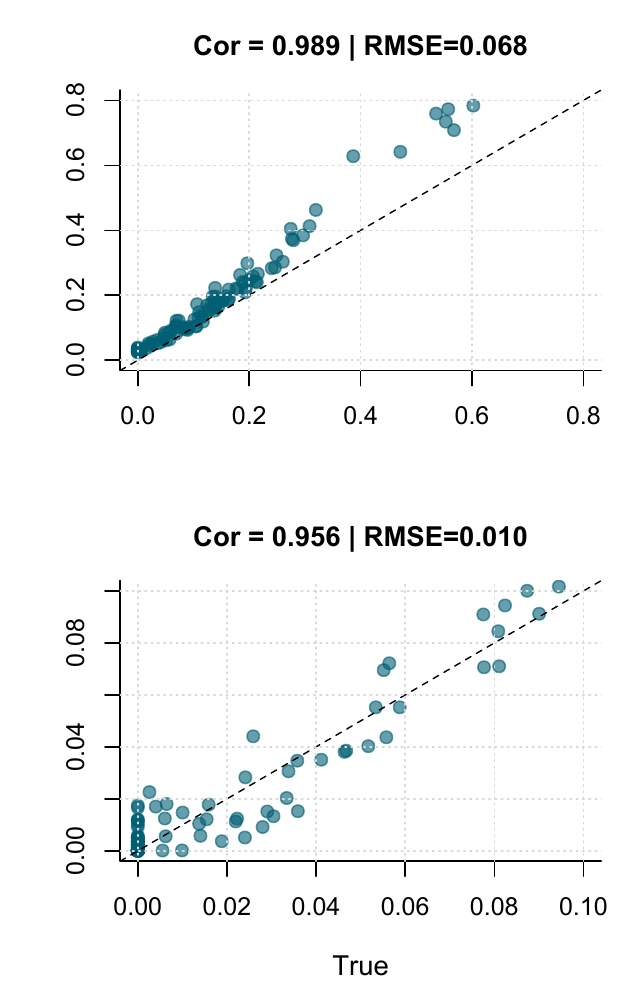}
    \caption{$\hat{\rho}_j^t = \hat{\rho}_j^l $}
    \label{fig:sim_cor_1rho}
    \end{subfigure}%
    \hfill
    \begin{subfigure}[t]{0.3\textwidth}
        \centering
        \includegraphics[width=0.95\linewidth]{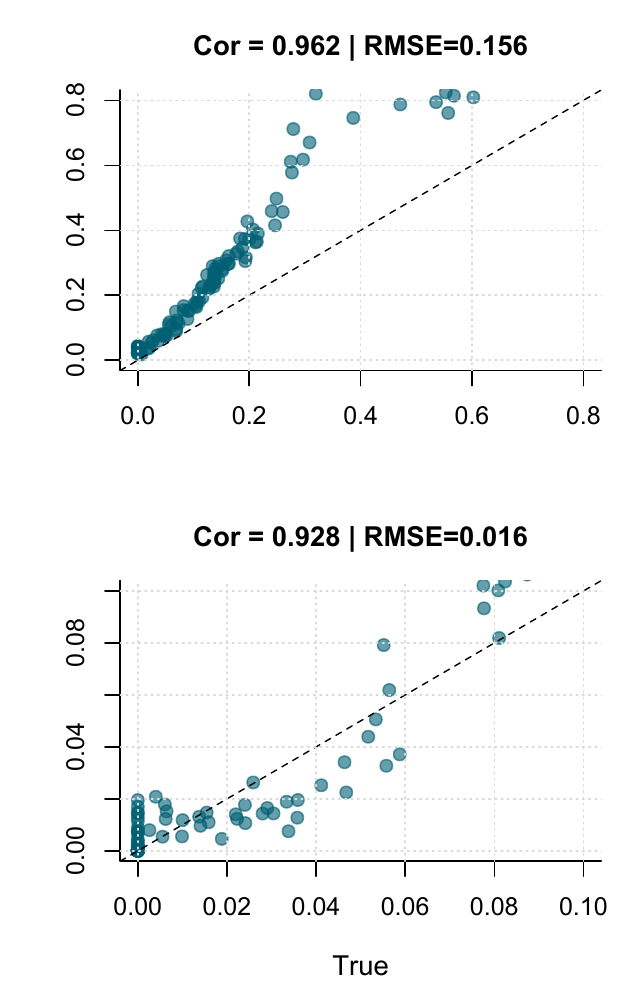}
    \caption{$\hat{\rho}_j^t = \hat{\rho}_j^l =0$}
    \label{fig:sim_cor_0rho}
    \end{subfigure}%
    \caption{Association of estimated and true tumour proportions when initialisation values of tumour means and variances are not exact.}
    \label{fig:sim_cor_mut3}
\end{figure*}

\begin{figure*}[!h]
    \centering
        \begin{subfigure}[t]{0.3\textwidth}
        \centering
        \includegraphics[width=0.95\linewidth]{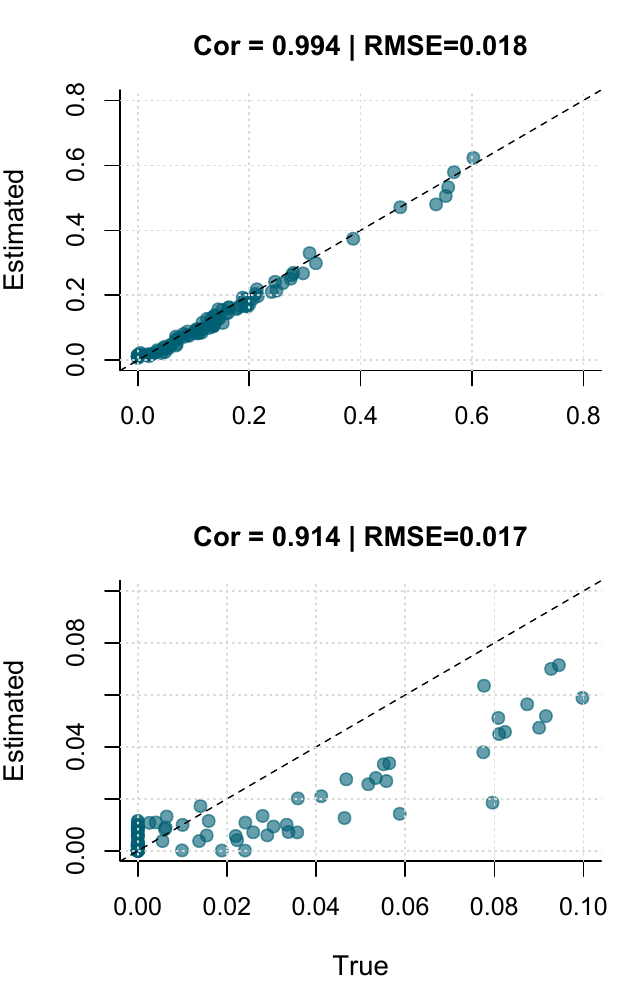}
    \caption{$\hat{\rho}_j^t \neq \hat{\rho}_j^l $}
    \label{fig:sim_cor_2rho_true0}
    \end{subfigure}%
    \hfill
    \begin{subfigure}[t]{0.3\textwidth}
        \centering
        \includegraphics[width=0.95\linewidth]{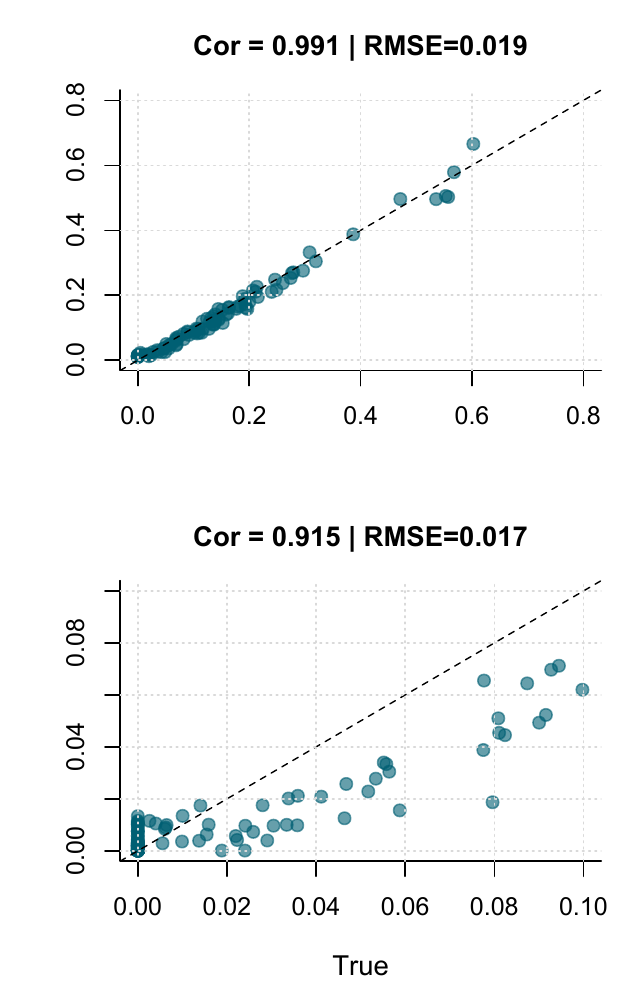}
    \caption{$\hat{\rho}_j^t = \hat{\rho}_j^l $}
    \label{fig:sim_cor_1rho_true0}
    \end{subfigure}%
    \hfill
    \begin{subfigure}[t]{0.3\textwidth}
        \centering
        \includegraphics[width=0.95\linewidth]{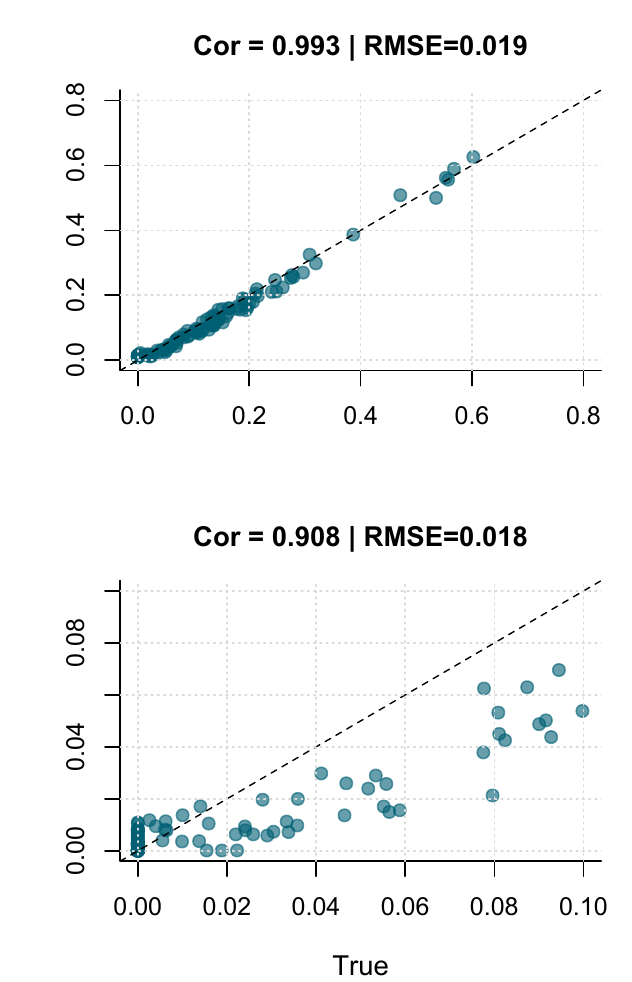}
    \caption{$\hat{\rho}_j^t = \hat{\rho}_j^l =0$}
    \label{fig:sim_cor_0rho_true0}
    \end{subfigure}%
    \caption{Association of estimated and true tumour proportions when initialisation values of background and tumour means and variances are not exact and the data are not correlated between measurements.}
    \label{fig:sim_cor_0rho_mut3_truerho0}
\end{figure*}

The association plots for the estimated tumour fractions under constant initial values are displayed in Figure \ref{fig:sim_cor_mut3}. Again, starting values for the tumour means were set to $\mu_j^t=3$ for all features. As a sanity check we repeated the analysis with data that with no correlation between measurements. The association results for the univariate data are plotted in Figure \ref{fig:sim_cor_0rho_mut3_truerho0}. As stated in the main text, these checks show how the bivariate approach performs better under constant initial values and does not result in meaningfully worse estimates when the data is not bivariate.

Given this constant starting values for the latent variables' distribution parameters, both bivariate specifications have a better performance estimating the tumour fractions compared to the univariate one when the data are correlated between the pre- and post-surgery measurements. When the data are actually univariate, the bivariate specification does not seem to hamper the estimation, as the correlation parameter is estimated to be close to zero within the first couple of iterations.

\section{Extended real data analysis: bivariate vs. univariate deconvolution} \label{app:ext_res}

\begin{figure*}[!h]
    \centering
    
    \begin{minipage}{0.46\linewidth}
        \centering
        \includegraphics[width=\linewidth]{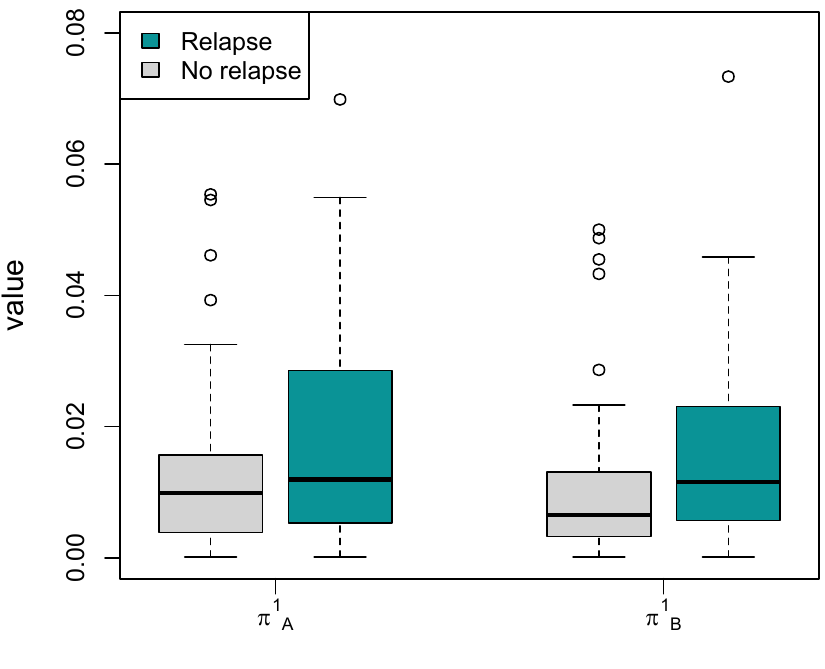}
        \caption{Estimated tumour proportions with respect to relapse status for the univariate ($\pi_A^1$) and the bivariate ($\pi_B^1$) specifications for all $112$ patients.}
        \label{fig:boxplot_app}
    \end{minipage}
    \hfill
    \begin{minipage}{0.48\linewidth}
        \centering
        \includegraphics[width=\linewidth]{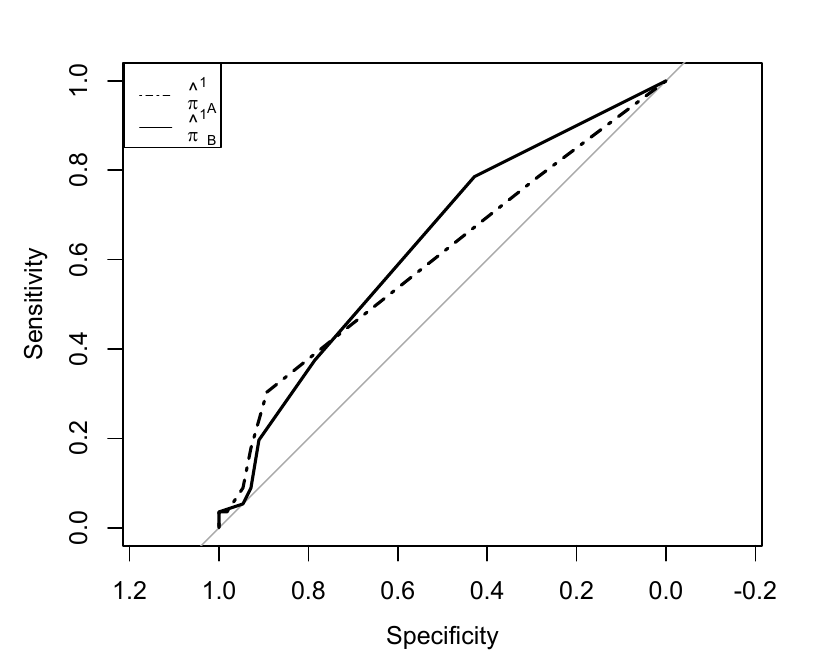}
        \caption{Estimated tumour proportions with respect to relapse status for the univariate ($\pi_A^1$) and the bivariate ($\pi_B^1$) specifications for all $112$ patients.}
        \label{fig:roc_app}
    \end{minipage}

\end{figure*}

In this section, we provide results for the estimated post-surgery tumour fractions under both the univariate ($\pi_A^1$) and the bivariate ($\pi_B^1$) deconvolution models, computed for all $112$ patients. These estimates are displayed in Figure \ref{fig:boxplot_app}. As stated in the main text, there no substantial differences between the distributions of the estimated tumour fractions from the subset of patients with available $\hat{\bpi}^1_{\text{VAF}}$, add those from the full cohort. 

The estimated tumour proportions under the bivariate model align closely with those presented in figure \ref{fig:boxplot}. Similarly, the performance metrics in Figure \ref{fig:roc_app} and in table \ref{tab:sumres_app} exhibit consistent patterns with the results discussed in the main text. This correspondence in the analysis between subsets under the bivariate model contrast to that of the univariate model. In particular, results for the latter show no significant distinction between the \textit{relapse} and the \textit{no relapse} groups, no significant link to said recurrence outcome and a lower AUC. Altogether, the bivariate deconvolution model showed a stronger and more robust link to the clinical outcomes compared to its univariate counterpart. 

\begin{table}[]
    \centering
\begin{tabular}{lccc}
\hline\hline
 & & \multicolumn{2}{c}{$\hat{\pi}^1$} \\ 
\cline{3-4}
 & &A & B  \\
\hline
Wilcoxon ($p$-value)  & & 0.227 & 0.012 \\
Fisher ($p$-value)  && 0.571 & 0.037  \\
Logistic ($\beta$)  && 0.286 & 0.871*  \\
AUC  && 0.6 & 0.63 \\
\hline\hline
\end{tabular} \vspace{1em}
\captionof{table}{Comparison between the post-surgery tumour fraction estimates for $\rho_j^b =\rho_j^t=0$, column A, and $\rho_j^b\neq\rho_j^t$, in column B for all $112$ patients.}
    \label{tab:sumres_app}
\end{table}

\begin{figure}[!h]
    \centering
    \begin{subfigure}[t]{0.45\textwidth}
        \centering
        \includegraphics[height=1.5in]{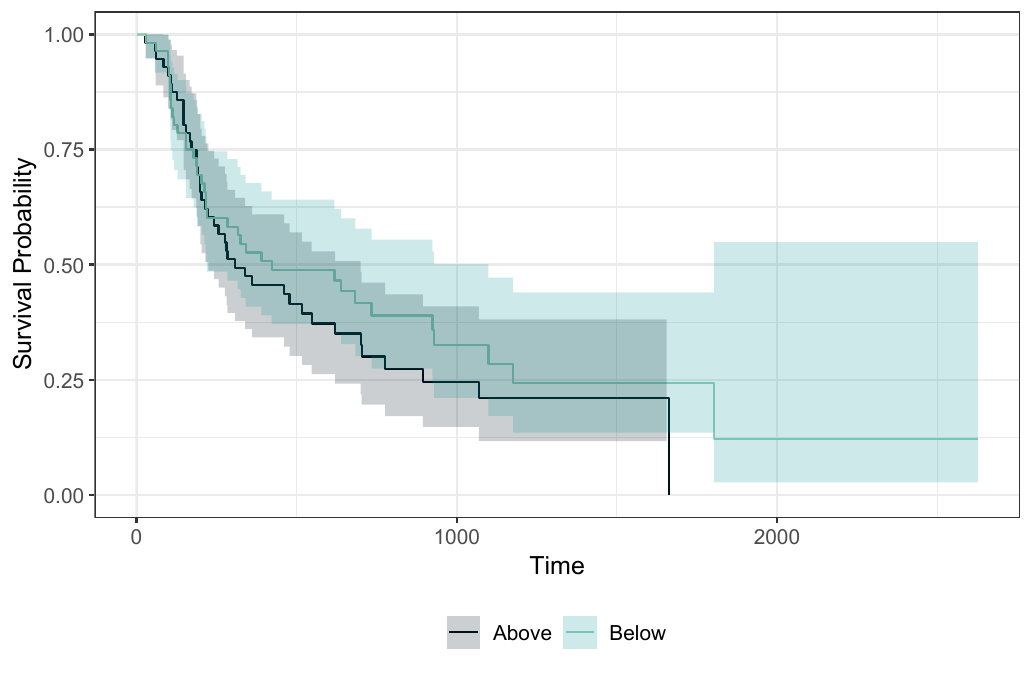}
        \caption{RFS curves for VAF values.}
    \end{subfigure}%
    ~ 
    \begin{subfigure}[t]{0.45\textwidth}
        \centering
        \includegraphics[height=1.5in]{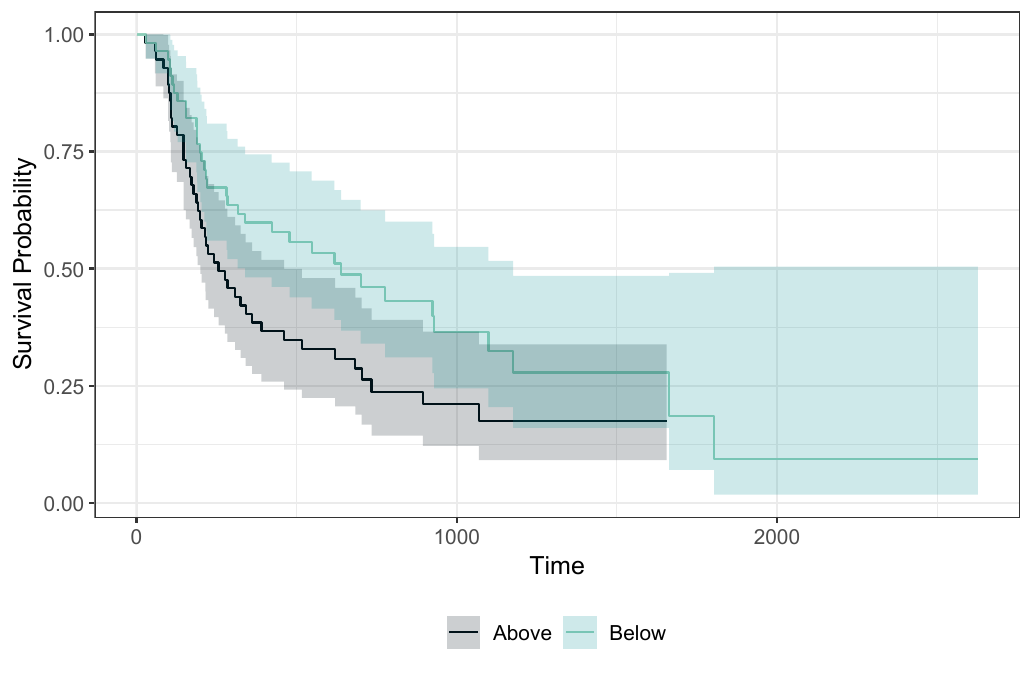}
        \caption{RFS curves for $\boldsymbol{\hat{\pi}}^1$.}
    \end{subfigure}
    \begin{subfigure}[t]{0.45\textwidth}
        \centering
        \includegraphics[height=1.5in]{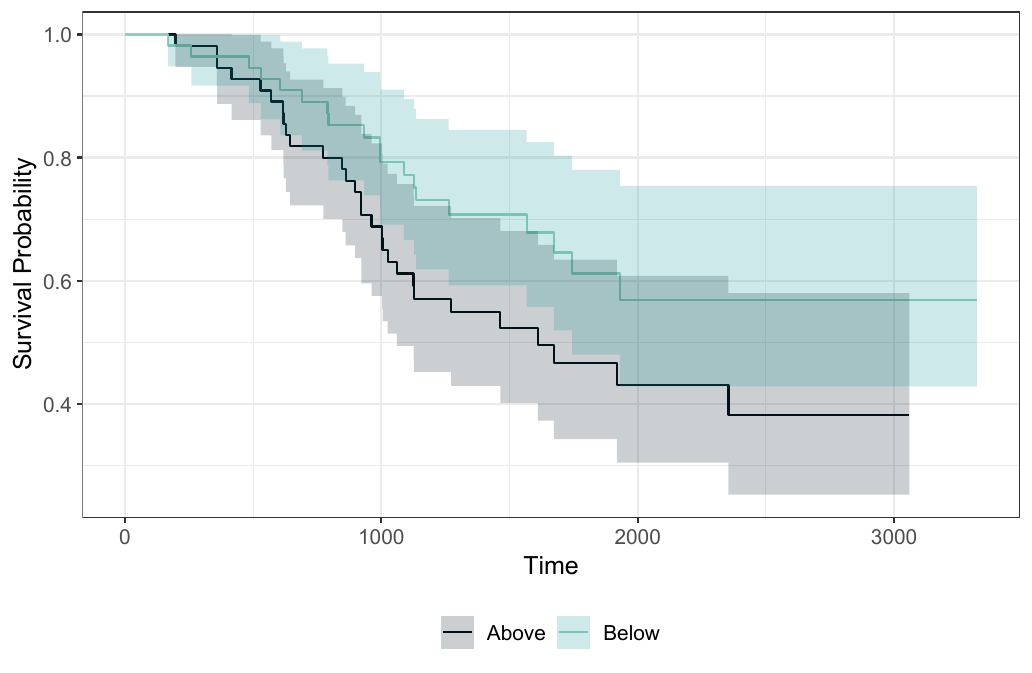}
        \caption{OS curves for VAF values.}
    \end{subfigure}
    ~ 
    \begin{subfigure}[t]{0.45\textwidth}
        \centering
        \includegraphics[height=1.5in]{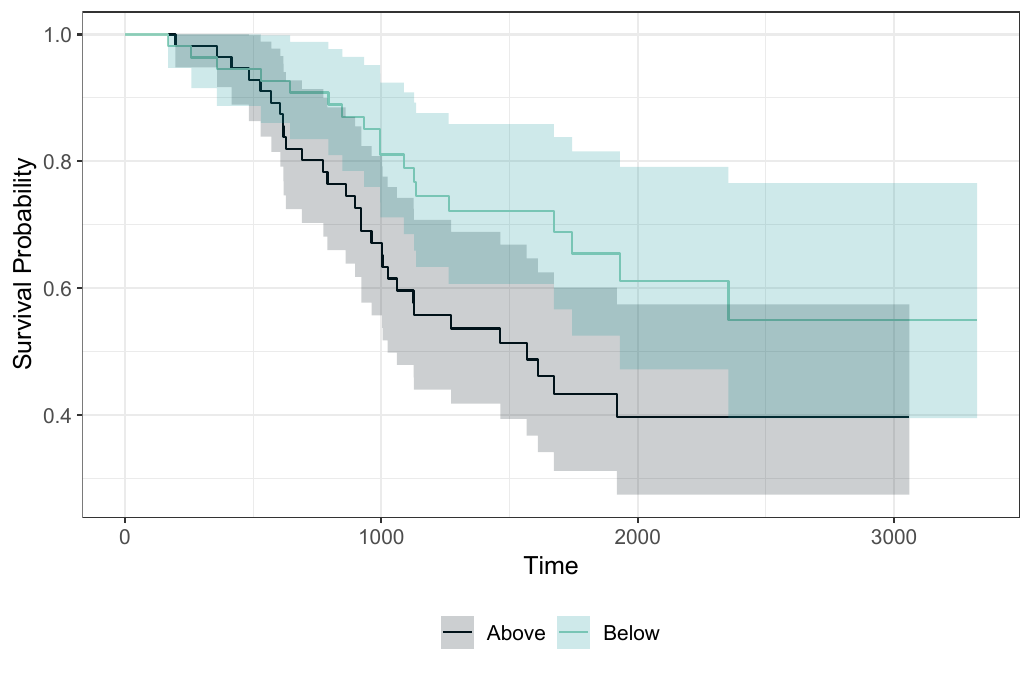}
        \caption{OS curves for $\boldsymbol{\hat{\pi}}^1$}
    \end{subfigure}
    \caption{Kaplan-Meier curves for OS, in days.}
    \label{app:kmc}
\end{figure}

\end{document}